\documentclass[prd,aps,colorBG,nofootinbib]{revtex4}
\usepackage[T2A]{fontenc}
\usepackage{amsmath,amssymb}
\usepackage{eucal}
\usepackage{color}
\usepackage[dviwindo]{graphicx}
\newcommand{\bs}{\begin{subequations}}
\newcommand{\es}{\end{subequations}}
\numberwithin{equation}{section}
\newcommand{\ben}{\begin{eqnarray}}
\newcommand{\een}{\end{eqnarray}}
\newcommand{\la}{\label}

\begin{document}

\title{Solutions of the Klein-Gordon equation on
 manifolds with variable geometry including dimensional reduction}

 \author{P. P.  Fiziev\\
 Dept. Theoretical Physics, Sofia University
 ``St. Kliment Ohridski", \\ 5 James Bourchier
 Blvd., 1164 Sofia, Bulgaria\\ and
 \\ BLTF, JINR, Dubna, 141980 Moscow Region, Russia \\ and
 \\ D.V. Shirkov, \\ BLTF, JINR, Dubna, 141980 Russia}

\begin{abstract}
 We develop  the recent proposal to use
 dimensional reduction from the four-dimensional space-time $D=(1+3)$
 to the variant with a smaller number of space dimensions $D=(1+d)$,
 $d < 3\,$ at sufficiently small distances to construct a renormalizable quantum field theory.
 We study the Klein-Gordon equation on a few toy
 examples ("educational toys") of a space-time with variable special geometry, including
 a transition to a dimensional reduction. The examples considered contain a combination
 of two regions with a simple geometry (two-dimensional cylindrical surfaces with different radii)
 connected by a transition region.
 The new technique  of transforming the study of solutions of the Klein-Gordon problem on
 a space with variable geometry into solution of a one-dimensional stationary Schr\"odinger-type equation with
 potential generated by this variation is useful. We draw the following conclusions:
 (1) The signal related to the degree of freedom specific to the higher-dimensional part
 does not penetrate into the smaller-dimensional part because of an inertial force inevitably
 arising in the transition region (this is the centrifugal force in our models).
 (2) The specific spectrum of scalar excitations resembles the spectrum of the real particles;
 it reflects the geometry of the transition region and represents its "fingerprints".
 (3) The parity violation due to the asymmetric character of the construction of our models
 could be related to violation of the CP symmetry. %


\end{abstract}
\sloppy
\maketitle

\section{Introduction}
 The main problem of standard quantum gravity with
 the classical Einsten-Hilbert action is related to the fact
 that the Newton constant $G_{\!{}_D},$ has a negative mass dimension
 $[G_{\!{}_D}]=[M^{2-D}]$ (in terms of a proper mass scale $M\,$),
 where $D=1+d$ is the space-time dimension. Quantum
 gravity is not perturbatively renormalizable for
 $D > 2$. The same is true for the electroweak theory
 without the Higgs field.
 This research is connected with \cite{Shirkov}, where a simple palliative was proposed
 for temporarily neutralizing of the nonrenormalizability problem.
 The point is to use dimensional reduction (also see \cite{Dejan10}) from the
 manifold with dimensionality $(1+3)\equiv 4$ to one with the smaller dimensionality
 $(1+d)$, $d< 3\,$ at sufficiently small distances (large momentum transfer).

 Our approach does not assume any modification of the concept
 of time. Instead, we have in mind some smooth enough
 reduction of spatial dimensions. As a result, the
 physical space is continuous, but may be not infinitely smooth manifold,
 and consists of parts with different topological dimension.
 Below, we use some toy models of space-time with variable geometry.
 To acquire some physical intuition and experience, we
 start with scalar wave solutions of the Klein-Gordon equation (KGE).

 We consider the KGE for a complex scalar
 field $\varphi(x)$ on a $(1+d)$-dim space-time with the
 signature $\{+,-,\dots,-\}$, $\mu,\nu=0,1\dots d$,
 ${x}=\{x^0, x^1,\dots x^d\}= \{x^0,\textbf{x}\}\,,$ where
 $x^0=t$ and $d\geq 1$ is an integer dimension of the
 space:
\ben
\la{KGE}
 \Box\varphi-M^2\varphi=0\,;\quad \Box= -{\frac{1}{
 \sqrt{|g|}}}\partial_\mu\left(\sqrt{|g|}g^{\mu \nu}
 \partial_\nu\right).\een
 Commonly, after quantization \cite{Pauli41} (see also Chap. 1 in \cite{BSh Book} and \cite{BSh Text}), a complex
 field $\varphi({x})\neq \overset{*}\varphi({x})$ corresponds
 to charged particles with electric charge $Q=\pm 1$ and
 (real) mass $M$ with current $
 j_\mu = i\left(\overset{*}\varphi \partial_\mu\varphi-
 \varphi\partial_\mu\overset{*}\varphi\right)\,$
  satisfying the continuity equation yielding the charge
 conservation law.

 The simplest possibility for choosing  a reference
 system is to use Gaussian coordinates
 in which the (1+d)-dimensional operator $\Box$
 (at least locally) takes the form
\ben\la{Box-Delta}
 \Box=-\partial^2_{tt} +\Delta_d,\quad \Delta_d={\frac{1}
 {\sqrt{|\gamma|}}}\partial_m\left(\sqrt{|\gamma|}
 \gamma^{m n}\partial_n\right)\,,\quad m,n=1,...,d.\een
 with the $d$-dimensional Laplacian $\Delta_d$ standardly defined by the
 $d$-dimensional metric $\gamma_{m n}$.
 From the very beginning we work in the
 framework of the $(1+d)$-formalism with a global time that
 is obligatory for constructing a correct physical
 picture and, especially, a quantum field theory.
 As a consequence,  the frequency
 $\omega$ of our solutions does not change on the parts with different topological dimension.

\section{The KGE in a cylindrical space geometry}

\subsection{The KGE in (1+2)-dimensions with cylindrical space symmetry}
We consider the KGE
\ben\la{WE}
\Box\varphi-M^2\varphi=-\partial^2_{tt} \varphi+\Delta_2
 \varphi- M^2\varphi=0 \een
 on a two-dimensional space manifold $\mathbb{M}^2_{\phi z}$ with
 cylindrical symmetry. In the Cartesian coordinates it is defined by
 a shape function $\rho(z)$ as a surface of rotation in
the  three-dimensional  Euclidean manifold $\mathbb{R}^{3}_{X^1 X^2 X^3}$:
 $$X^1=\rho(z)\cos\phi, \quad X^2=\rho(z)\sin\phi,\,\,\,X^3 = z.$$
 The restriction of the  three-dimensional  interval
 $(dL)^2=(dX^1)^2+(dX^2)^2+(dX^3)^2$ on the two-dimensional  manifold
 $\mathbb{M}^2_{\phi z}$ is written as
\ben\la{2Dmetric}
(dl)^2=\gamma_{mn}dx^mdx^n=\rho^2(z)\,d\phi^2+\left(1+
 {\rho^\prime}^2 \right)dz^2,\quad\rho^\prime=d\rho/dz.\een
The Laplacian in explicit form
\ben
\Delta_2=
{\frac 1 {\rho^2}}\left(\partial^2_{\phi\phi}+{\frac \rho
 {\sqrt{1+{\rho^\prime}^2}}} \partial_z{\frac \rho
 {\sqrt{1+{\rho^\prime}^2}}}\partial_z\right)
\la{lapalcian}
\een
admits the separation of the variables $\varphi(t,\phi,z)=T(t)
 \Phi(\phi) Z(z)$, yielding a system of ordinary differential
 equations (ODEs). Two of them are simple:
$T^{\prime\prime}+\omega^2 T=0,\,\,\,\Rightarrow\,\,\, T(t)=
 e^{-i\omega t}$ and
 $\Phi^{\prime\prime}+m^2\Phi=0,\quad\Rightarrow\,\,\,
 \Phi(\phi)=e^{im\phi},\,\,\,m=0, \pm1,\pm2,\dots$\ . The only
 nontrivial equation is for the function $Z(z)$,
\ben\label{Z_z}
{\frac 1 {\rho\sqrt{1+{\rho^\prime}^2}}} \partial_z\left(
 \!{\frac \rho {\sqrt{1+{\rho^\prime}^2}}}\partial_z Z\!\right)+
 \left(\!\omega^2-M^2-{\frac {m^2} {\rho(z)^2}}\!\right)Z&=0,
 \quad &\Rightarrow\,\,\, Z(z)=Z(z;\omega,m). \la{ode:c}\een
 It contains the remarkable term $m^2/\rho(z)^2$ with the form
 of the potential energy of centrifugal force acting for $m\neq 0$.
 We note that this
 term presents the simplest example of an inertial
 force with a transparent physical meaning.
 Such forces are inevitable for motion in a curved space-times.
 They arise in the transition regions between the parts of space with
 different topological dimensions.
The hope therefore arises that we can learn something about the possible physical effects
of the inertial forces acting in the class of curved apace-times
with variable dimensions described above.

 \subsection{Transformation of the problem to solution of
 Schr\"odinger-type ODE.} A useful mathematical property for studying solutions
 of  Eq.\eqref{Z_z} is contained in the following theorem.\bigskip

 {\bf \emph{Theorem.}} \emph{Stationary equation \eqref{Z_z} can be
 transformed into the stationary Schr\"odinger-type equation}
\ben\la{Schrodinger}
U^{\prime\prime}(u)+\big(E-V(u)\big)U(u)=0.
\een

We note that the possibility of reducing solutions of the KGE to solutions of a
Schr\"odinger-type equation arises because of the separation of variables in the KGE.
Moreover, it uses  the fact that
 any two-dimensional metric $\gamma_{mn}$ is conformally flat.
 Transforming the coordinate $z$ to $u$ and the shape
 function $\rho(z)$ to $\varrho(u)=\rho(z(u))$,
\ben\la{uz_int}
 z \mapsto u:\quad u(z)=\int\sqrt{1+{\rho^\prime(z)}^2}{\frac{dz}
 {\rho(z)}};\hskip 1.truecm u \mapsto z:\quad z(u)=
 \int\sqrt{\varrho(u)^2-\varrho'(u)^2}\,du\,,\een
 one obtains the Laplacian \eqref{lapalcian} in the form
 $\Delta_2=\varrho^{-2}\left(\partial^2_{\phi\phi}+
 \partial^2_{uu}\right)$, which shows that the function $\varrho(u)^2$ in our problem
 is precisely the two-dimensional conformal factor. In terms of variable $u$, Eq. \eqref{ode:c}
 takes the form \eqref{Schrodinger}:
\ben\la{eqU}
U^{\prime\prime}(u)+\big(-\left(M^2-\omega^2\right)\varrho(u)^2  -m^2\big)U(u)=0,
\een
 with identification
 $E=0,\,V(u)=\left(M^2-\omega^2\right)\varrho(u)^2+m^2$,
 $Z(z)=U(u(z))$.

 Studying the KGE on the considered curved manifolds thus
 reduces to soluving the Schr\"odinger-type
 equation with the potential $V(u)$ defined by the
 geometry. The broad literature devoted to the properties of such equations and their solutions
 for concrete potentials $V(u)$ (see, e.g., \cite{Regge_dAlfaro},
 \cite{Fluge}) can now be used for our problem.
We also note that
there is no additive spectral parameter in Eq. \eqref{eqU} like the usual quantum
energy $E$ in \eqref{Schrodinger} contained in the term $E-V(u)$.
Instead, the specific spectral parameter $\omega$ of our problem
appears in the  factor $M^2-\omega^2$ in the potential function $V(u)$.
This detail somewhat subtly modifies the well known spectra of quantum problems
presented in the literature. In particular, some solutions of Eq. \eqref{Schrodinger}
that are below the ground state and have no physical meaning in quantum mechanics case
"surface into the physical region" in the KGE context here, as it were,
because of the change of the ground state
when $M^2-\omega^2$ changes it sign.

This result can be considered a realization of the old idea by Hertz \cite{Hertz}
about a geometrical description of the physical forces, but now in the wave language.
It can also be applied in the theory of wave guides in acoustics and
optics, for radio-waves, and in other branches of wave physics.

\subsection{The scalar wave equation in (1+1)-dimensional
 spacetime}
 For a one-dimensional case with the coordinate $z$ and  Laplacian $\Delta_1=\partial^2_{zz}$,
 the wave function has the simple form
 \begin{subequations}\label{phi1:a,b}
 \begin{eqnarray}
\varphi_1^{\,Q\pm}(t,z)&=\text{const}\times e^{-i\omega t}e^{\pm i\sqrt{\omega^2-M^2} z}:\,\,\,
Q&=+1 \Rightarrow \text{for particles},\label{phi1:a}\\
\overset{*}\varphi_1\!{}^{Q\pm}(t,z)&=\text{const}\times e^{i\overset{*}\omega t}e^{\mp i\sqrt{\overset{*}\omega{}^2-M^2} z}:\,\,\,
Q&=-1 \Rightarrow \text{for antiparticles}.\label{phi1:b}
\end{eqnarray}
\end{subequations}
The additional superscript $\pm$ denotes the sign of the momentum $p_z=\pm k=\pm \sqrt{\omega^2-M^2}$,
or complex conjugated momentum $\overset{*}p_z=\pm \overset{*}k=\pm \sqrt{\overset{*}\omega^2-M^2}$.
With the conventional $\Re(\omega)\geq 0$
we have the standard relativistic combination
$-i(\omega t-p_z z)=-ipx$ in the solutions for particles and
$i(\omega^* t-p_z^* z)=(-ipx)^*$ in the conjugate solutions for antiparticles.

\section{Some explicit examples}

\subsection{Two cylinders of constant radii $R$ and $r<R$,
 connected by a part of cone}
\begin{figure}[htbp]
\vskip -1.2truecm
\centering
\begin{minipage}{12.cm}
\hskip -8.truecm
\includegraphics[width=1.6truecm, viewport=4 1 100 300]
 {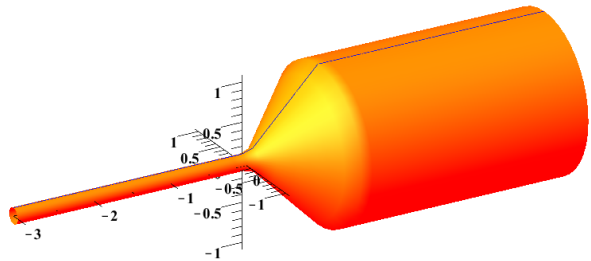} \vskip -.8truecm
\caption{\small Surface of two cylinders of radii $R$ and
 $r<R$ connected by a part of  cone}
\label{Fig1}
\end{minipage}
\end{figure}
\begin{figure}[htbp]
\vskip -.5truecm
\centering
\begin{minipage}{12.5cm}
\hskip -7.8truecm
\includegraphics[width=1.5truecm, viewport=4 1 100 300]
 {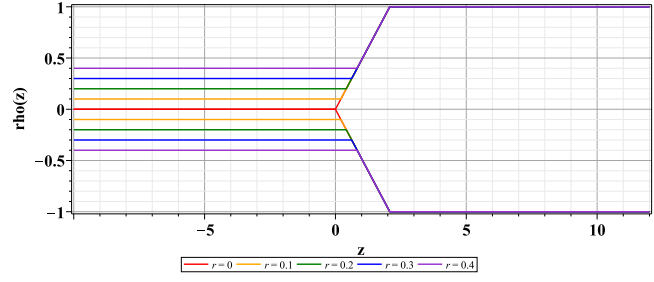}
 \vskip -.3truecm
 \caption{\small The section of continuous 
 manifolds consisting of two cylinders with radii $R=1$ and $r=0,
 0.1, 0.2, 0.3, 0.4$ connected by part of a cone with the divergence angle
 $\alpha=\pi/7$}
\label{Fig2}
\end{minipage}
\end{figure}
\begin{figure}[!htb]
\vskip 1.3truecm
\begin{center}
\begin{minipage}{8.1truecm}
\vskip .truecm
\hskip -5.truecm
\includegraphics[width=2.5truecm, viewport=4 1 200 200]
 {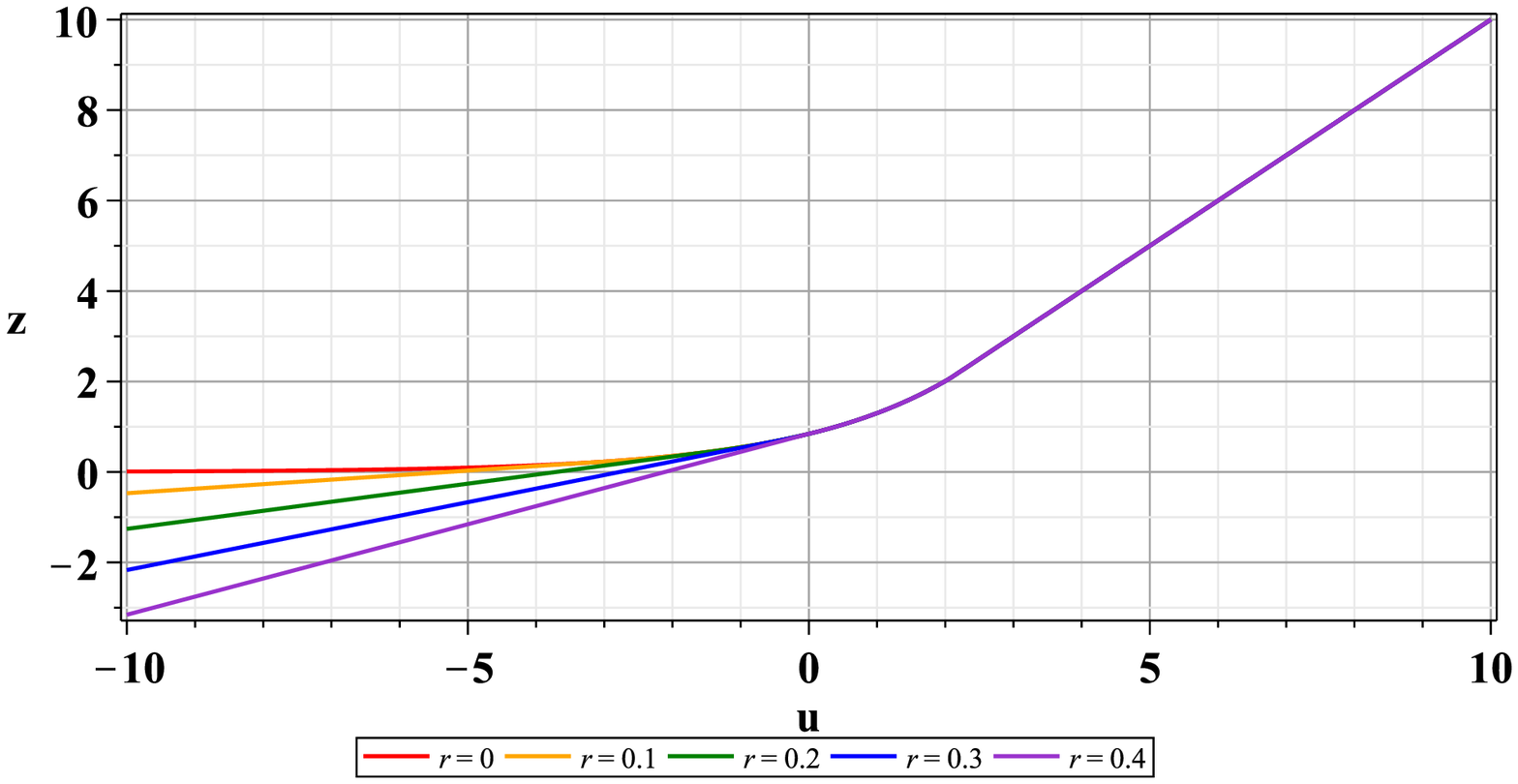}
\vskip -.1truecm
\caption{\small The limit $r\to 0$ for the function
 $z(u;R,r,\alpha)$}
\label{lim_z_r}
\end{minipage}
\hspace{.5truecm}%
\begin{minipage}{8.1truecm}
\vskip -.57truecm
\hskip -5.3truecm
 \includegraphics[width=2.65truecm, viewport=4 230 200 200]
 {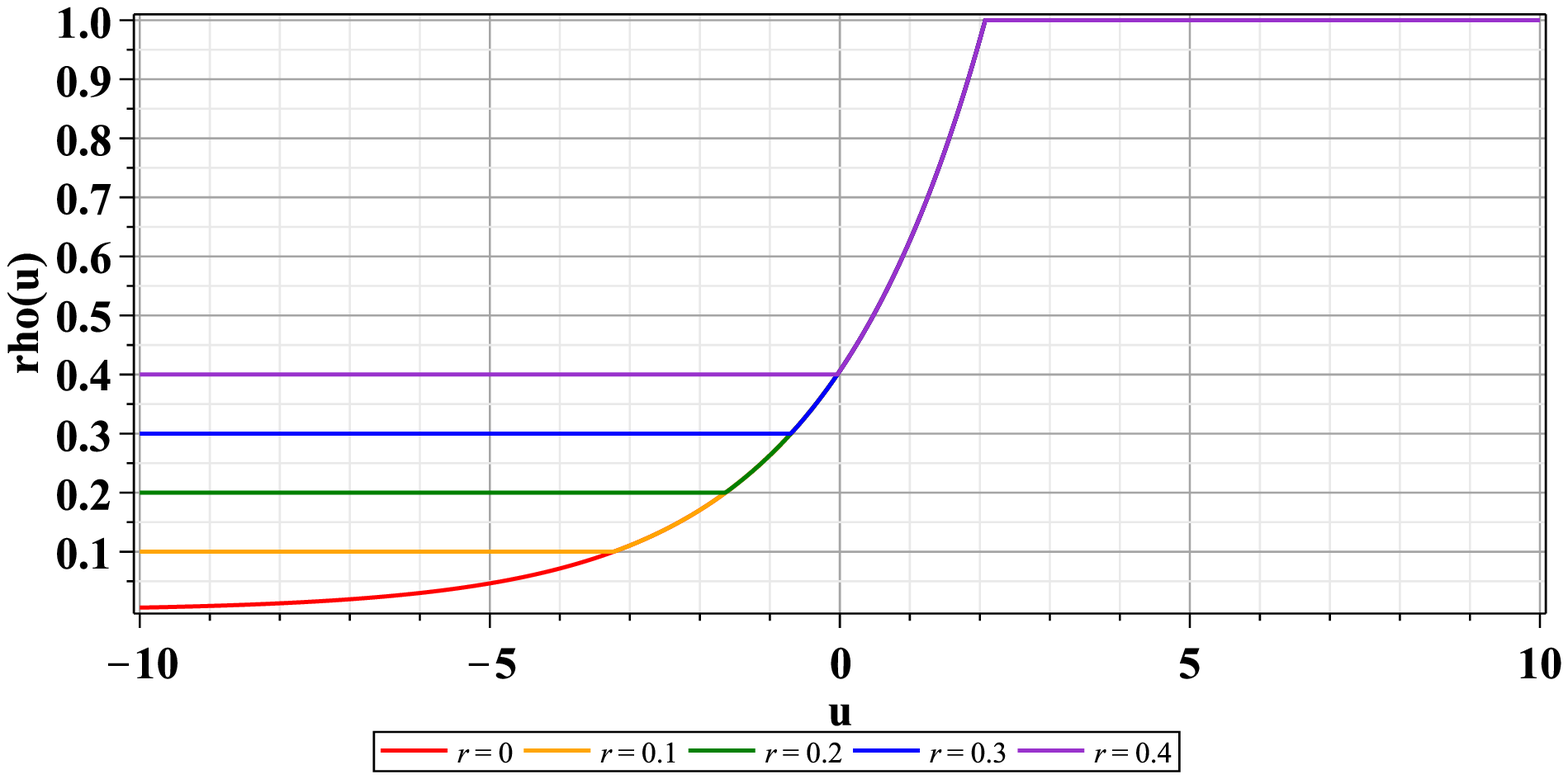}
 \vskip 2.9truecm
 \caption{\small The limit $r\to 0$ for the function
 $\varrho(u;R,r,\alpha)$}
\label{lim_rho_r}
\end{minipage}
\end{center}
\end{figure}

 We turn to the simple case of the two-dimensioanl space \cite{Shirkov}: the
 surface of two cylinders of radii $R$ and $r<R$ continuously connected
 by part of a cone (see Fig.\ref{Fig1}). Let the symmetry axis  be the
 horizontal axis $Oz$ with the origin at the cone vertex.
 The shape function $\rho(z)$ (see Fig. \ref{Fig2}) then has the form
 \ben\la{rho_z}
\rho(z)=\left\{\begin{array}{cccc}
 R=\text{const}&:\hskip .2truecm \text{for}\quad z&\in& [z_R,
 +\infty), \cr
 z\tan\alpha&: \hskip .2truecm \text{for}\,\,\, z &\in&
 [z_r,z_R],\cr
 r=\text{const}&:\hskip .2truecm \text{for}\,\,\, z&\in&
 (-\infty,z_r].
 \end{array}\right.\een
 Here $\alpha\in (0,\pi/2)$ is half of the (fixed) cone vertex angle,
 $z_{\!{}_R}=R\cot\alpha>0$,
 and $z_r=r\cot\alpha>0$.
The functions $z(u)$ and $\varrho(u)$ (see Figs. \ref{lim_z_r} and \ref{lim_rho_r}) are obtained from \eqref{uz_int}:
\ben
z(u;R,r,\alpha)=\left\{\begin{array}{ccc}
R\,u : \hskip .3truecm &\text{for}&\,\,\, u\geq u_R, \cr
R\,u_R\exp(u\sin\alpha-\cos\alpha) : \hskip .3truecm &\text{for}&\,\,\, u \in [u_r,u_R],\cr
r\big(u+\ln(R/r)/\sin\alpha\big) : \hskip .3truecm &\text{for}&\,\,\, u\leq u_r,
\end{array}\right.
\la{z_u}
\een
\ben
\varrho(u;R,r,\alpha)=\left\{\begin{array}{ccc}
R=\text{const}: \hskip .3truecm &\text{for}&\,\,\, u\geq u_R, \cr
R\exp(u\sin\alpha-\cos\alpha) : \hskip .3truecm &\text{for}&\,\,\, u \in [u_r,u_R],\cr
r=\text{const}: \hskip .3truecm &\text{for}&\,\,\, u\leq u_r,
\end{array}\right.
\la{rho_u}
\een
where $u_R=\cot\alpha$ and $u_r=\cot\alpha -
 \ln(R/r)/\sin\alpha$.

For waves on the surface of a cylinder with an arbitrary radius $\rho$ we have
the dispersion relation:
\ben
k_{\rho}=\sqrt{\omega^2-M^2-{\frac {m^2}{\rho^2}}}.
\la{dispersion}
\een

 Under normalization:
$\varphi^{Q\pm}_{\omega,m}=(a/\sqrt{2})e^{-iQ\omega t}\,
 e^{i m \phi}e^{\pm k_{\rho}z}$.
 the complex solutions of KGE whith the charge $Q=\pm 1$
 and mass $M$ on the cylinder with radius $\rho$
 generate a conserved current with the components
\ben
j^t= Q\,|a|^2{\mathfrak{Re}(\omega)},\,\,\,
j^\phi=m|a|^2,\,\,\,
j^z=\pm |a|^2\mathfrak{Re}\left(k_{\rho}\right).
\la{curl}
\een
As can be seen, for $m\neq 0$ the wave rotates on the surface of the cylinder because $J^\phi\neq 0$  in Eq. \eqref{curl}.
This rotation of the scalar wave generates a centrifugal force.
It becomes clear that for $m\neq 0$ in the limit $\rho\to \rho_{crit}=|m|/\sqrt{\omega^2-M^2}$
the height of the centrifugal potential barrier increase without bounded and stops all physical signals.
In contrast, no such obstacle exists for a nonrotational movement along the axis $Oz$ with $m=0$.
The physical signals can propagate without obstacles on the common physical degree of freedom
of the two parts of the physical space with different topological dimension.

Using the standard continuity conditions for the function $Z(z)$ and its derivative $Z^\prime(z)$  we obtain
(see the notation in Appendix A)
\ben
\overrightarrow{Z}(z;\omega,m;R,r,\alpha)=-\Theta^{+}_r e^{i k_r z_r}\left\{\begin{array}{ccc}
{\frac 2 \pi}{\frac{2ik_r z_r}\Delta}e^{i k_{\!{}_R}(z-z_{\!{}_R})}:\,\,\,z\geq z_{\!{}_R},\cr
\cr
{\frac{2ik_r z_r}{\Delta}}
\left(Y^{-}_R\, J_\nu\left(k_c z\right)-J^{-}_R\, Y_\nu\left(k_c z\right)\right):\,\,\,\, z_r\leq z\leq z_{\!{}_R},\cr
\cr
-e^{i k_r(z-z_r)}+{\frac{\Delta_{11}}{\Delta}}e^{-i k_r(z-z_r)}:\,\,\,\,z\leq z_r,
\end{array}\right.
\la{Z_right}
\een
for the waves going to the right from $z=-\infty$ to $z=+\infty$ experiencing reflection on the junction
and
\ben
\overleftarrow{Z}(z;\omega,m;R,r,\alpha)=-\Theta^{-}_{\!{}_R}
 e^{-i k_{\!{}_R} z_{\!{}_R}}\left\{\begin{array}{ccc}
-e^{-i k_{\!{}_R}(z-z_{\!{}_R})}+{\frac{\Delta_{22}}{\Delta}}\,
 e^{i k_{\!{}_R}(z-z_{\!{}_R})}:\,\,\,z\geq z_{\!{}_R},\cr
\cr
{\frac{2i k_{\!{}_R} z_{\!{}_R}}{\Delta}}
\big(Y^{+}_r\, J_\nu\left(k_c z\right)-J^{+}_r\, Y_\nu
 \left(k_c z\right)\big):\,\,\,\, z_r\leq z\leq z_{\!{}_R},\cr
\cr
{\frac 2 \pi}{\frac{2i k_{\!{}_R} z_{\!{}_R}}{\Delta}}
 e^{-i k_r(z-z_r)}:\,\,\,\,z\leq z_r.
\end{array}\right.
\la{Z_left}\een
for waves going to the left from $z=+\infty$ to $z=- \infty$.
 We consider the limit $r\to +0\,,\,R=\text{const}$,
 $\alpha=\text{const}$ in relation \eqref{rho_z} when the two-dimensional
 cylinder of radius $r$ with $z\leq z_r$ transforms into the one-dimensional
 manifold, an infinitely thin thread stretched along the negative part of the axis $Oz$. From
a geometrical standpoint, this is a very singular limit (see
 Figs. \ref{lim_z_r} and \ref{lim_rho_r}).
 In this limit $z_{\!{}_R}=R\cot\alpha$ does not change,
 $z_r\to 0$ and the third piece of the functions
 $z(u;R,r,\alpha)$ and $\varrho(u;R,r,\alpha)$ disappears
 because $u_r$ goes to $-\infty$. As a result, the shape
 function $\varrho(u;R,r,\alpha)$, or the corresponding
 conformal factor $\varrho(u;R,r,\alpha)^2$ is unsuitable
 for describing the one-dimensional part of the emergent continuous
 (but not smooth) manifold with variable dimension because the
 infinite interval $u\in (-\infty,\infty)$ is mapped only to
 the semi-infinite interval $z\in [0,\infty)$ when $r=0$.

\subsubsection{The states of the continuum spectrum}

 If $J^{-}_R\neq 0$, then for real  frequencies
 $\omega\geq M$ we obtain
\ben
\mathbf{S}(\omega,m;R,\alpha)=\lim_{r\to 0}\mathbf{S}(\omega,m;R,r,\alpha)=
-\Theta_{\!R}^{-}e^{-i 2 k_{\!{}_R} z_{\!{}_R}}\,
\left({{J^{+}_R}/{J^{-}_R}}\right)
\left(\begin{array}{cc}
0 & 0 \\
0 & 1
\end{array}\right).
\la{Sr0}
\een

Several important consequences in the case $m\neq 0$ become obvious.

i) In the limit $r\to +0$ the transition coefficients are $|S_{12}|^2=|S_{21}|^2\equiv 0$. Hence,
the communication by wave signals with an azimuthal $m\neq 0$
between the domains with different dimension is impossible .

ii) In fact, the limit $r\to +0$ of the solutions incoming from $z=-\infty$ is trivial for any $m\neq 0$:
$$\overrightarrow{Z}(z;\omega,m;R,\alpha)=\lim\limits_{r\to 0}\overrightarrow{Z}(z;\omega,m;R,r,\alpha) \equiv 0$$
everywhere because the factor $\Theta^{+}_r=\Theta\left(\omega - \sqrt{M^2+m^2/r^2}\right)\equiv 0$ for $r<\sqrt{\omega^2-M^2}/|m|$
in \eqref{Z_right}.
In addition $S_{11}\to 0$, $S_{12}\to 0$.

iii) In the two-dimensional domain, the
 modulus of the reflection coefficient for real frequencies $\omega$ is $|S_{22}|\equiv 1$, i.e.,
 we have a total reflection on the cone of the waves incoming
 from $z=+\infty$, accompanied by a change of the phase of the
 scattered wave, according to Eq. \eqref{Sr0}. Hence, now we
 obtain the nontrivial solutions with $m\neq 0$:
\ben
\overleftarrow{Z}(z;\omega,m;R,\alpha)\!=\!\lim\limits_{r\to 0}\overleftarrow{Z}(z;\omega,m;R,r,\alpha)
\!=\!-\Theta_{\!R}^{-}e^{-i k_{\!{}_R} z_{\!{}_R}}\left\{\begin{array}{ccc}
\!-e^{-i k_{\!{}_R} \left(z-z_{\!{}_R}\right)}
\!+\!\left({{J^{+}_R}/{J^{-}_R}}\right)\,
e^{i k_{\!{}_R}\left(z-z_{\!{}_R}\right)},\,\,\,z\geq z_{\!{}_R},\cr
\cr
\left(2ik_{\!{}_R} z_{\!{}_R}/J^{-}_R\right)\,J_\nu\left(k_c z\right),\,\,\,
0\leq z\leq z_{\!{}_R},\cr
\cr
0,\,\,\,z\leq 0.
\end{array}\right.
\la{Solution_m_r0_left}
\een
We note the following features of these solutions:

1. They are regular and finite everywhere, including the singular point $z=0$ where they vanish.

2. They do not penetrate nor propagate into the one-dimensional part of the space.

3. There are no solutions with nonnegative real $\omega<\sqrt{M^2+m^2/R^2}$
because the factor $\Theta_{\!R}^{-}$ vanishes.
This is obviously a correct physical result because
the group velocity of the waves vanishes  for $\omega=\sqrt{M^2+m^2/R^2}$.
In addition, the wavelength becomes infinite in this case.

4. The waves propagating on the cone junction have a complicated continuous spectrum of momenta $p_z$.
 The spectrum can be obtained by the Fourier transform of solution
 \eqref{Solution_m_r0_left} with respect to the variable $z$.
 This spectrum is a sort of \ "fingerprint"  characterizing the geometry
 of the transition region (a cone in this case).
 By studying this spectrum,
 we can hope to reconstruct the geometry of the junction, et
 least to some extent. An analogous problem was first posed in acoustics
 as early as 1877 by Lord Rayleigh, then substantially advanced in 1911 by Hermann Weyl
 and later by many others (see the recent review article
 \cite{Weyl} and the references therein).

\subsubsection{The resonant states}
\begin{figure}[!htb]
\vskip 2.truecm
\begin{center}
\begin{minipage}{8.1truecm}
\vskip .2truecm
\hskip -4.5truecm
\includegraphics[width=2.3truecm, viewport=4 1 200 200]
 {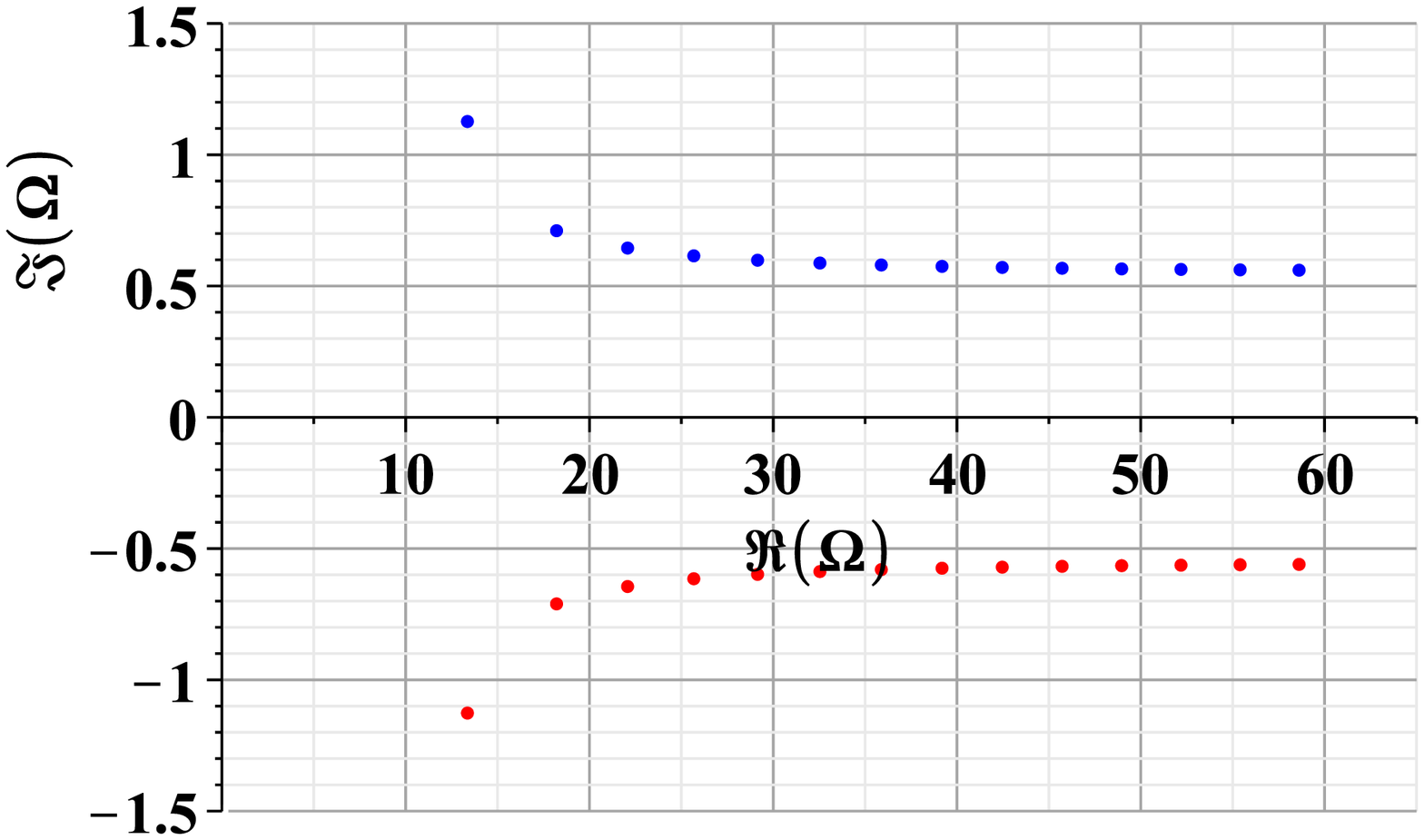}
\vskip -.2truecm
\caption{\small The complex spectrum $\omega^{{}_Q}_{n,m}$,
 $n=0,1,2,3,\dots$
 for $m=10$ and $\alpha=\pi/3$ for mass $M=0$ and
 $R=\sin(\alpha)$. Red points are particles and blue points are antiparticles}
\label{f61}
\end{minipage}
%
\hspace{.5truecm}%
\begin{minipage}{8.1truecm}
\hskip -5.truecm
\includegraphics[width=2.2truecm, viewport=4 230 200 200]
 {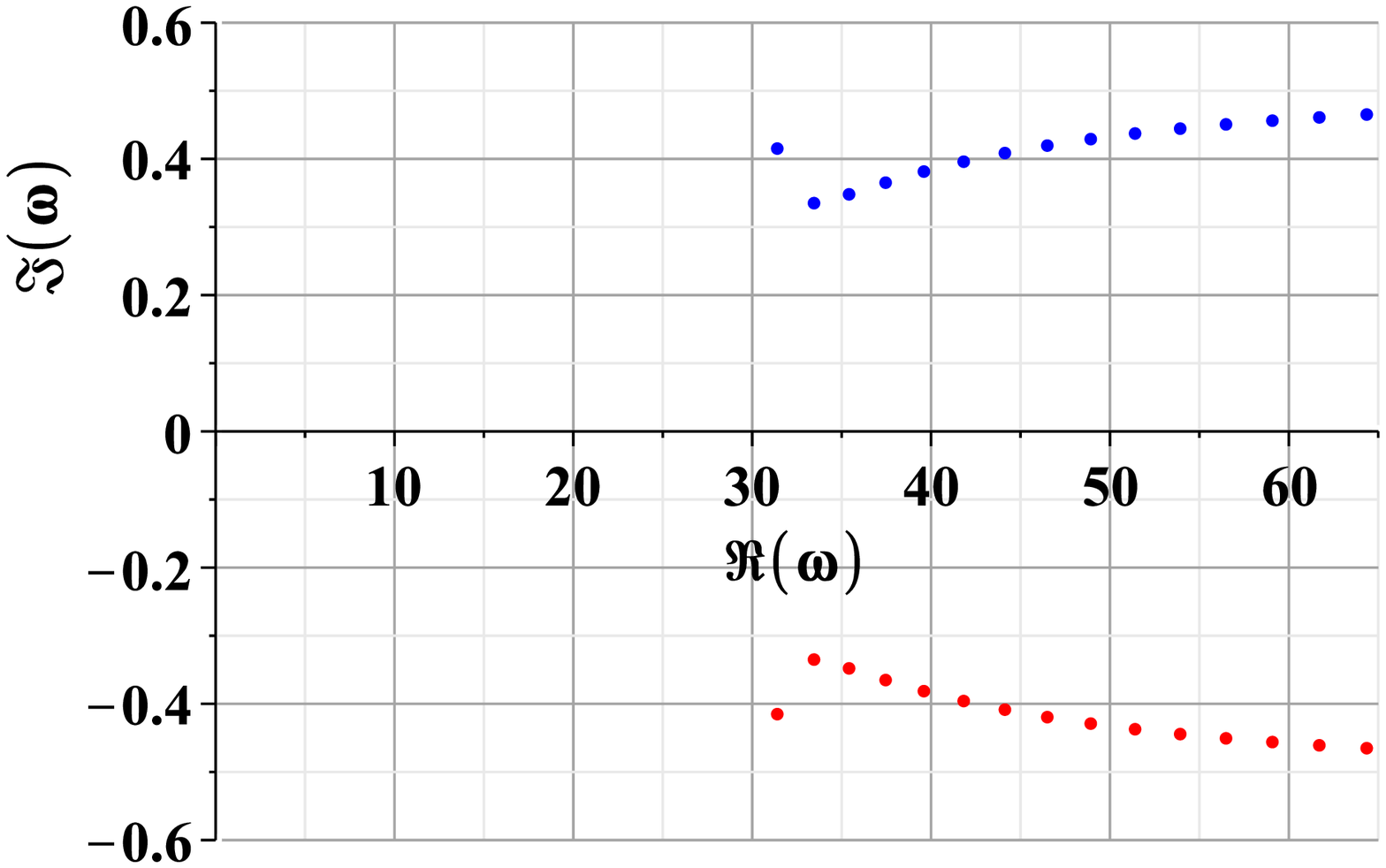}
\vskip 2.3truecm
\caption{\small The complex spectrum $\omega^{{}_Q}_{n,m}$,
 $n=0,1,2,3,\dots$ for $m=10$ and $\alpha=\pi/3$ for mass
 $M=28.85$ and $R=\sin(\alpha)$.Red points are particles and blue points are antiparticles}
\label{f71}
\end{minipage}
\end{center}
\end{figure}

 Up to now,  we have worked far from the poles of the scattering matrix $\mathbf{S}$
 defined as zeros of the denominator in \eqref{Sr0} or
 Eq. \eqref{Solution_m_r0_left}. For some $m\neq 0$ we
 consider a discrete (infinite) sequence of frequencies
 $\omega_{n,m},\,n=0,1,\dots$, satisfying the  condition $J^{-}_R=0$.
 Using the dimensionless variable $\Omega=k_c z_R$ we write
 the spectral condition in the form ${{\Omega J^\prime_
 \nu(\Omega)}/{J_\nu(\Omega)}}=\cos\alpha \sqrt{\nu^2-\Omega^2}$.
 A typical example of the absence of real roots and a representative
 sequence of complex roots is shown in Figs. \ref{f61} and
 \ref{f71}. As can be seen, there is an infinite number
 of complex resonant frequencies
\begin{equation}
\omega^Q_{n,m}=\sqrt{M^2
+\frac{\sin^2\alpha}{R^2}\bigl(\Omega^Q_{n,m}\bigr)^2},
\label{omega_Omega}
\end{equation}
 which correspond to the complex roots $\Omega_{n,m}$
 and depend to some extent on the mass $M$ of the scalar
 field (see  Figs. \ref{f61}, \ref{f71}). The spectrum
 $\omega^{{}_Q}_{n,m}$ characterizes the geometry of the
 conic form of the junction and the continuous but not
 smooth transition between the two cylinders.
%

 Using the asymptotic expansion of the Bessel functions
 $J_\nu(\Omega)$ for large $|\Omega|\gg \nu=|m|/\sin\alpha$ one
 obtains the asymptotic form of the complex roots:
\ben\la{Omega_asymptotic}
\Omega^{{}_Q}_{n,m}\approx\pi\left(n+{\frac{|m|}{2\sin\alpha}}-
 {\frac 1 4}\right) - i\, Q\,\text{artanh}(\cos\alpha),\quad
 n=0,1,2,\dots;\, m=0,\pm 1,\pm2,\dots.\een
 The choice of the sign of the imaginary part of
 $\Omega^{{}_Q}_{n,m}$ in \eqref{Omega_asymptotic} ensures the exponential
 decay of the field excitations as $t\to +\infty$ for both particles
 $Q=+1$ and antiparticles $Q=-1$.

The simple formula \eqref{Omega_asymptotic} to some
 extent reflects the influence of the  geometry of the junction
 from the two-dimensional space to the one-dimensional space
 on the physics of resonant excitations of the Klein-Gordon field.

 When the divergence angle $\alpha$ of the cone tends to the upper limit $\alpha\to \pi/2$ and $n\gg |m|$, we have
 the relation
$$
\omega^Q_{n,m}\to
\sqrt{M^2+\frac{\pi^2}{R^2}\biggl(n+\frac{|m|}{2}-\frac{1}{4}\biggr)^2}\,.
$$
In this limit the halflive of the resonant excitation increases without bounded,
$\tau_{n=0,m}=1/|\Im(\omega^{{}_Q}_{n=0,m})|\to \infty$, i.e.,
the resonant excitations are preserved for a very long time.

 In the opposite, case when the angle $\alpha\to 0$,
 approximation
 \eqref{Omega_asymptotic} itself is unapplicable because
 $|\Omega^{{}_Q}_{n,m}|\sim \pi\nu\to\infty$ and the
 condition $|\Omega^{{}_Q}_{n,m}|\gg\nu$ is not satisfied
 as needed, because $|\Omega^{{}_Q}_{n,m}|/\nu\sim \pi>1$ but
 $\pi \not\gg 1$. The corresponding more accurate estimates
 can be found in \cite{DVPF}.

 Hence, we observe that the change in the topological dimension of space yields a new
 physical effect. Effective mass-spectrum \eqref{omega_Omega} of
 excitations  and their lifetime depend
 on  the junction geometry  between the two-dimensional
 and one-dimensional spaces.

Using the simplest normalization we write down two equivalent representations of the relativistic  resonance wave functions:
\begin{subequations}\label{3.12:a,b}
\begin{align}
Z_{\mathrm{res}}(z;\omega_{n,m},m;R,\alpha)&=
\begin{cases}
e^{ik_R(z-z_R)}, &z\geq z_R,
\\
Z_{\mathrm{res,c}}(z_;\omega_{n,m},m;R,\alpha), &z_r\leq z\leq z_R, \quad \text{where} \label{3.12:a}\\
0, &z\leq z_r,
\end{cases}\\
Z_{\mathrm{res,c}}(z_;\omega_{n,m},m;R,\alpha)&=
\dfrac{\pi}{2}\,Y^-_RJ_\nu(k_\mathrm{c}z)\equiv\dfrac{2ik_Rz_R}{J^+_R}\,J_\nu(k_\mathrm{c}z),\label{3.12:b}
\end{align}
\end{subequations}
obtaining an identity for Bessel functions and their derivatives as a byproduct.

\subsection{A conelike junction with smooth transition to the cylinder}
We consider the special case of the function $\varrho(u)=
R\exp(p u)/\left(1+\exp(p u)\right)$\footnote{Other
examples of analytic solutions of the KGE on two-dimensional manifolds
with cylindrical symmetry can be seen in \cite{DVPF}.}.
In the variables $\rho$ and $z$, the corresponding form function of the surface $\rho(z)$ is given by
\begin{equation}
\frac{z(\rho)}{R}
=\frac{1}{p}\ln\frac{1+\sqrt{1-p^2(1-\rho/R)^2}}{1-\rho/R}
-\frac{1}{p}\sqrt{1-p^2\biggl(1-\frac{\rho}{R}\biggr)^2}
+\mathrm{const}.
\label{zu_rho_u_pp1}
\end{equation}
\begin{figure}[htbp]
\vskip -1.1truecm
\centering
\begin{minipage}{13.cm}
\hskip -7.7truecm
\includegraphics[width=1.6truecm, viewport=4 1 100 300] {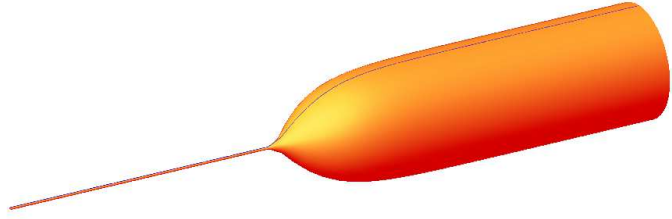}
\vskip -1.truecm
\caption{\small A conelike junction with a smooth transition to the cylinder}
\label{Con_Cyl_smooth}
\end{minipage}
\end{figure}
implicitly defying a two-dimensional surface (See Fig. \ref{Con_Cyl_smooth}.)
The possible forms of its sections are shown in Fig. \ref{p_qp_1}.
At the vertex, as before, this is a conelike surface with the angle $\alpha=\arctan\left(p/\sqrt{1-p^2}\right)$.
Therefore $p=\sin\alpha$.
\begin{figure}[htbp]
\vskip -.9truecm
\centering
\begin{minipage}{9.3cm}
\hskip -7.truecm
\includegraphics[width=1.4truecm, viewport=4 1 100 300] {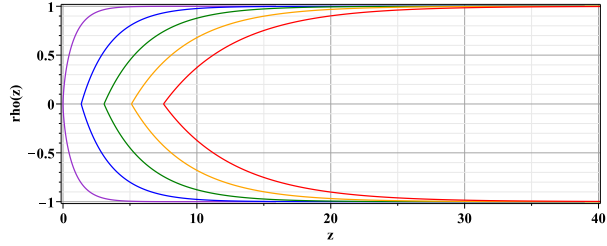}
\vskip -.3truecm
\caption{\small The sections of the two-dimensional surface $\rho(z)$ for $R=1$ and
$p=\sin\alpha=0.185$-red, $0.23$-gold, $0.30$-green, $0.43$-blue, $1.00$-violet}
\label{p_qp_1}
\end{minipage}
\end{figure}

For $E=0$ the Schr\"odinger-like equation ~\eqref{Schrodinger} with effective potential
$V(u)=V_0/\left(1+e^{-pu} \right)^2+m^2$, where $V_0=R^2(M^2-\omega^2)$, has  two
independent solutions:
\begin{align}
U^\pm(u)&=A_\pm(u)e^{\pm mu}= \notag\\
&=\frac{\Gamma(c_\pm)\Gamma(b_\pm-a_\pm)}{
\Gamma(b_\pm)\Gamma(c_\pm-a_\pm)}B_\pm(u)e^{-u\sqrt{V_0+m^2}}+
\frac{\Gamma(c_\pm)\Gamma(a_\pm-b_\pm)}{
\Gamma(a_\pm)\Gamma(c_\pm-b_\pm)}C_\pm(u)e^{+u\sqrt{V_0+m^2}},
\label{3.14}
\end{align}
(see the notation in Appendix B).
In addition we have $A_\pm(u)\to1$ for $u\to-\infty$, and $B_\pm(u)$,
$C_\pm(u)\to1$ for $u\to+\infty$.

If we substitute $|m|$ instead of $\pm m$ in all places in Eq. \eqref{3.14},
then we obtain a solution $\stackrel{\leftarrow}{U}(u;\omega,m;R,a)$,
analogous to the solution \eqref{Solution_m_r0_left} and also vanishing at the cone vertex.

The poles of the $S$-matrix are defined by the equations $b_+=-n$, $n=0,1,2,\dots$, or $c_+-a_+=-n$, $n=0,1,2,\dots$\,.
As a result, we obtain the sequence of frequencies:
\begin{equation}
\omega_{n,m}
=\sqrt{M^2+\frac{m^2}{R^2} -
\frac{m^2-p^2/4}{R^2}\check{Z}^2
\biggl(\frac{(2n+1)p/2+|m|}{\sqrt{m^2-p^2/4}}
\biggr)},\qquad n=0,1,2,\dotsc,
\label{3.15}
\end{equation}
where we use the Zhukovsky function $\check{Z}(x)=(x+1/x)/2$.
It can be seen that for  $M>0$ between  frequencies~\eqref{3.15}, we have
a finite number of real positive frequencies, for  $n\in[0,n_{\max}]$, where
$n_{\max}=[\mathcal{N}]$ is the integer part of the number
$$
\mathcal{N}=\sqrt{\biggl(\frac{MR}{p}\biggr)^2+\biggl(\frac{m}{p}\biggr)^2}-
\frac{|m|}{p}+
\sqrt{\biggl(\frac{MR}{p}\biggr)^2+\biggl(\frac{1}{2}\biggr)^2}-\frac12>0.
$$
Because the real frequencies are less then $\sqrt{M^2+m^2/R^2}$, the corresponding momenta
$\sqrt{\omega_{n,m}^2-M^2-m^2/R^2}$ are purely imaginary.
Such solutions correspond to series of bound states with different $n$ and $m$,
which decay exponentially at space infinity.

For $n>n_{\max}$, formula \eqref{3.15} gives an infinite series of purely imaginary frequencies.


\section{Summary and outlook}

 We generalize our results.\\

 1. We have proved a useful theorem that relates studying the Klein-Gordon equation on spaces with variable
 geometry to solving a Schr\"odinger-type equation with an effective potential generated by the
 geometry variation. This result is based on separation of variables in the KGE and on the fact that two-dimensional spaces
 are conformally flat. We showed that in the case of space dimension $d=2$, the conformal factor of the metric enters
 the effectice potential in the Schr\"{o}dinger-type equation because of the corresponding changes of variables.
 The generalization to space dimensions $d>2$ was studied in \cite{Fiziev2010}.\\

 2. As a corollary, we conclude that the obtained
  field excitation spectra could serve as
 "fingerprints" of the form $\rho(z)\,$ of the junction region. The obtained nontrivial spectra
 for scalar excitations qualitatively resemble some actual spectra of resonances of elementary particles.\\

 3. Signals related to the degrees of freedom of only the higher-dimensional part of space do not propagate
 freely trough the junction region (in particular, these
 signals do not penetrate into the smaller-dimensional part).
 In our toy models,
 the reason is the centrifugal force acting on the
 solutions with nonzero angular momenta. In the limit
 case of dimensional reduction, this force grows infinitely
 and blocks all the solutions but one with azimuthal number $m=0\,.$
 We can say that signals that penetrate from higher dimensions into lower ones are only
 the signals associated with common degrees of freedom.\\

 Studying variable geometry in more realistic
($d=3$) spaces is definitely interesting for further research.
 This case is close to problems in wave physics (acoustics,
wave guides,etc.).

 The parity violation (P-violation) due to the
 asymmetry of the space geometry is obvious. Hence,
 spaces with variable geometry can perhaps provide a simple
 basis for describing the real situation with the
 C, P and T properties of the particles.

 The second result listed above  suggests a new idea.
 We can try to reproduce the observed
 spectra of elementary particles using an appropriate
 geometry of the junction region between parts of the
 space-time with different topological dimension.

\begin{acknowledgments}
The authors are pleased to thank I. Aref'eva and O. Teryaev for the useful discussions
and L. Simeonov for the help in calculations.
One of the aithors (D.V.Sh.) expresses his gratitude to S. Rastopchina
for the permanent support over the last 60 years.

This research was supported in part by the
Program for Supporting Leading Scientific Schools (Grant No. NSh-3810.2010.2),
the Russian Foundation for basic Research (Grant Nos. 08-01-00686 and 11-01-00182)
and the Bulgarian National Scientific Fund
(Contracts Nos. DO-1-872, DO-1-895 and DO-02-136).
\end{acknowledgments}

\appendix

\section{Brief notation for the formulas in Sec. 3}
In the problem with a conelike junction domain, we use the standard notation for the Bessel functions $J_\nu(x)$ and $Y_\nu(x)$.
Sometimes only $J_\nu(x)$ are called the Bessel functions and
$Y_\nu(x)$ are called the Neumann functions ($N_\nu(x)\equiv Y_\nu(x)$) \cite{SpecialFunctionsA,SpecialFunctionsB}.

The wave numbers along the $z$-axis on the corresponding cylinders are
$k_{\!{}_R}=\sqrt{\omega^2-M^2-m^2/R^2}$ $\big(\Re(k_{\!{}_R})\geq 0\big)$ and
$k_r=\sqrt{\omega^2-M^2-m^2/r^2}$ $\big(\Re(k_r)\geq 0\big)$.
We use the notation $k_c=\sqrt{\omega^2-M^2}/\cos\alpha$ on the cone.
We note that the wave number of the one-dimensional waves with $m=0$
propagating on the axis $Oz$ (see Eqs. \eqref{phi1:a,b})
is just the projection $k=k_c\cos\alpha=\sqrt{\omega^2-M^2}$ $\big(\Re(k)\geq 0\big)$.

We use the brief notation
\ben
J^{\pm}_\nu(x,\alpha)=\left(x{\frac{d}{dx}}\pm i\varkappa_\nu(x,\alpha)\right)J_\nu(x),\,\,\,
Y^{\pm}_\nu(x,\alpha)=\left(x{\frac{d}{dx}}\pm i\varkappa_\nu(x,\alpha)\right)Y_\nu(x),
\la{JYpm_x}
\een
\ben
\varkappa_\nu(x,\alpha)=\cos\alpha\sqrt{x^2-\nu^2},\,\,\,
\varkappa_R=\varkappa_\nu(k_c z_{\!{}_R},\alpha)=k_{\!{}_R}z_{\!{}_R},\,\,\,
\varkappa_r=\varkappa_\nu(k_c z_r,\alpha)=k_r z_r,
\la{varkappa}
\een
\ben
J^\pm_R =J^{\pm}_\nu(k_c z_{\!{}_R},\alpha),\,\,\,J^\pm_r=J^{\pm}_\nu(k_c z_r,\alpha),\,\,\,
Y^\pm_R =Y^{\pm}_\nu(k_c z_{\!{}_R},\alpha),\,\,\,Y^\pm_r=Y^{\pm}_\nu(k_c z_r,\alpha),
\la{JYpm}
\een
\ben
\Delta=J^{-}_R Y^{+}_r-J^{+}_r Y^{-}_R,\,\,\,\Delta_{11}=J^{-}_R Y^{-}_r-J^{-}_r Y^{-}_R,\,\,\,\Delta_{22}=J^{+}_R Y^{+}_r-J^{+}_r Y^{+}_R.
\hskip .truecm
\la{Delta_}
\een
This notation is useful for a compact description of the S-matrix in formulas \eqref{Z_right}-\eqref{Solution_m_r0_left}, and \eqref{3.12:a,b}.

\section{Brief notation in the formula (III.14) }
In formula \eqref{3.14} we use the brief notation
\begin{align*}
&a_\pm=\frac{1}{2}+\frac{1}{p}
\biggl(\sqrt{V_0+m^2}
-\sqrt{\frac{p^2}{4}+V_0}\pm m\biggr),
\\
&b_\pm=\frac{1}{2}+\frac{1}{p}
\biggl(-\sqrt{V_0+m^2}
-\sqrt{\frac{p^2}{4}+V_0}\pm m\biggr),\qquad
c_\pm=1\pm i\frac{2}{p}m,
\\
&A_\pm(u)=(1+e^{pu})^{1/2-\sqrt{1/4+V_0/p^2}}
{}_2F_1\bigl(a_\pm,b_\pm;c_\pm;-e^{pu}\bigr),\\
&B_\pm(u)=(1+e^{-pu})^{1/2-\sqrt{1/4+V_0/p^2}}
{}_2F_1\bigl(a_\pm,a_\pm-c_\pm+1;
a_\pm-b_\pm+1; -e^{-pu}\bigr),\\
&C_\pm(u)=(1+e^{-pu})^{1/2-\sqrt{1/4+V_0/p^2}}
{}_2F_1\bigl(b_\pm,b_\pm-c_\pm+1;
b_\pm-a_\pm+1; -e^{-pu}\bigr).
\end{align*}



\begin{thebibliography}{9}
%
\bibitem{Shirkov}  D.V. Shirkov, (2010) {\em Coupling running through
 the Looking-Glass of dimensional Reduction}, Particles and Nuclei (PEPAN), Letters, {\bf 7}, No 6(162); arXiv:1004.1510.

\bibitem{Dejan10} Luis Anchordoqui, et al., ``Vanishing
   Dimensions and Planar Events at the LHC.'' [hep-ph:1003.5914]


\bibitem{Pauli41} W. Pauli, (1941) {\em Relativistic Field Theories of
 Elementary Particles} Rev. Mod. Phys. {\bf 13}, 203-232; see also Ch. 1
  in the monograph \cite{BSh Book} and in the text-book \cite{BSh Text}.

\bibitem{BSh Book} N.N. Bogoliubov and D.V. Shirkov, (1980) {\em Introduction to the Theory of Quantized
                              Fields}, 3rd edition, Wiley-Interscience, N.Y.

\bibitem{BSh Text}  N.N. Bogoliubov and D.V. Shirkov, (1983) {\em Quantum Fields}, Benjamin/Cummings Publ., Reading.

\bibitem{Regge_dAlfaro} V. De Alfaro, T. Regge, (1965) {\em Potential Scattering}, North-Holland Publ. Comp., Amsterdam.

\bibitem{Fluge} S. Fl\"{u}gge (1971) {\em Practical Quantum Mechanics I}, Springer, Berlin, Heidelberg, N.Y.

\bibitem{Hertz} Heinrich R. Hertz {\em Die Principien der Mechanik im neuem Zusammennhange dargestellt}, Gesamellte Werke, Leipzig, 1910.

\bibitem{Weyl} Wolfgang Arendt, Robin Nittka, Wolfgang Peter, Frank Steiner (2009).
               {\em Weyl's Law: Spectral Properties of the Laplacian in Mathematics and Physics}
               in Mathematical Analysis of Evolution, Information, and Complexity.
               Edited by Wolfgang Arendt and Wolfgang P. Schleich,
               WILEY-VCH Verlag GmbH \& Co. KGaA,Weinheim

\bibitem{DVPF} Shirkov D. V., Fiziev P. P., {\em Amusing properties of Klein-Gordon solutions on manifolds with variable geometry},
               talk given at the  International Workshop "Bogoliubov Readings", September 22-25, JINR, Dubna, 2010. http://tcpa.uni-sofia.bg/index.php?n=7

\bibitem{Fiziev2010}  P.~P.~Fiziev {\em Partially Compact $\mathbf{(1+d)}$-dim Riemannian Space-Times
       which Admit Dimensional Reduction to Any Lower Space Dimension and the Klein-Gordon Equation}, arXiv:1012.3520 [math-ph].

\bibitem{SpecialFunctionsA} Abramowitz, Milton; Irene A. Stegun (1964). {\em Handbook of Mathematical Functions}, Nat. Bureau of Standards,
                           Appl. Math. Ser.

\bibitem{SpecialFunctionsB} Gradshteyn, I. S.; I. M. Ryzhik (2007). {\em Table of Integrals, Series, and Products}, Academic press.

\end{thebibliography}
\end{document}